\documentclass[pdflatex,sn-apa]{sn-jnl}% APA Reference Style
\usepackage{graphicx}%
\usepackage{multirow}%
\usepackage{amsmath,amssymb,amsfonts}%
\usepackage{amsthm}%
\usepackage{mathrsfs}%
\usepackage[title]{appendix}%
\usepackage{xcolor}%
\usepackage{textcomp}%
\usepackage{manyfoot}%
\usepackage{booktabs}%
\usepackage{algorithm}%
\usepackage{algorithmicx}%
\usepackage{algpseudocode}%
\usepackage{listings}%
\usepackage{array}%
\usepackage{makecell}%
\usepackage{float} 
\theoremstyle{thmstyleone}%
\theoremstyle{thmstyletwo}%

\theoremstyle{thmstylethree}%

\begin{document}

\title[Article Title]{Beyond Correctness: Evaluating and Improving LLM Feedback in Statistical Education}

\author*[1]{\fnm{Niklas} \sur{Ippisch}}\email{n.ippisch@campus.lmu.de}

\author[1]{\fnm{Markus} \sur{Herklotz}}
\equalcont{These authors contributed equally to this work.}

\author[1,2]{\fnm{Anna-Carolina} \sur{Haensch}}
\equalcont{These authors contributed equally to this work.}

\author[3]{\fnm{Carsten} \sur{Schwemmer}}
\equalcont{These authors contributed equally to this work.}

\affil*[1]{\orgdiv{Social Data Science and AI Lab}, \orgname{LMU Munich}, \orgaddress{\city{Munich}, \country{Germany}}}

\affil[2]{\orgdiv{University of Maryland College Park}, \orgaddress{\state{MD}, \country{USA}}}

\affil[3]{\orgdiv{Department of Sociology - Professorship of Computational Social Sciences}, \orgname{LMU Munich}, \orgaddress{\city{Munich}, \country{Germany}}}

\abstract{Large language models (LLMs) have been proposed as scalable tools to address the gap between the importance of individualized written feedback and the practical challenges of providing it at scale. However, concerns persist regarding the accuracy, depth, and pedagogical value of their feedback responses. The present study investigates the extent to which LLMs can generate feedback that aligns with educational theory and compares techniques to improve their performance. Using mock in-class exam data from two consecutive years of an introductory statistics course at LMU Munich, we evaluated GPT-generated feedback against an established but expanded pedagogical framework. Four enhancement methods were compared in a highly standardized setting, making meaningful comparisons possible: Using a state-of-the-art model, zero-shot prompting, few-shot prompting, and supervised fine-tuning using Low-Rank Adaptation (LoRA). Results show that while all LLM setups reliably provided correctness judgments and explanations, their ability to deliver contextual feedback and suggestions on how students can monitor and regulate their own learning remained limited. Among the tested methods, zero-shot prompting achieved the strongest balance between quality and cost, while fine-tuning required substantially more resources without yielding clear advantages. For educators, this suggests that carefully designed prompts can substantially improve the usefulness of LLM feedback, making it a promising tool, particularly in large introductory courses where students would otherwise receive little or no written feedback.}

\keywords{Automated Feedback Generation, Large Language Models, Prompt Engineering, Fine-Tuning, Feedback Quality Evaluation}

\maketitle

\section{Introduction}\label{sec1}

“The positive effect of feedback on students’ performance and learning is no longer disputed,” claim Lipnevich and Pandero (\citeyear[p.~1]{lipnevich_review_2021}) based on their study of 14 different feedback frameworks and their impact. But although providing individual and timely feedback to students is highly desirable in an educational context, it can pose a significant challenge to educators \citep{li_am_2023}. This is particularly true for large university courses with several hundred students. Teaching personnel may therefore find themselves in a situation where giving detailed feedback is simply not feasible.

Large Language Models (LLMs) have been proposed for possibly bridging that gap: Following the introduction of OpenAI’s\footnote{https://openai.com} ChatGPT, the use of LLMs in educational contexts has increased considerably \citep{jeon_large_2023}. Even though there are certainly advantages, such as scalability and accessibility, there are also concerns, both from students \citep{kayali_investigation_2023} and teaching personnel \citep{jeon_large_2023}, especially regarding potential errors and hallucinations. Those concerns are valid as research indicates that erroneous feedback can have significant negative effects on students \citep{li_am_2023}: Particularly, feedback that states that a wrong answer by a student is correct has negative impacts on students, as such errors are more difficult for students to detect \citep{li_am_2023}. Errors, however, are not the only possible problem when it comes to LLM-generated feedback. Furthermore, the quality of feedback can be low, which poses a problem, since the quality of feedback is a critical factor influencing the impact of feedback on students \citep{hattie_visible_2009}.

\begin{figure}
    \centering    \includegraphics[width=1\linewidth]{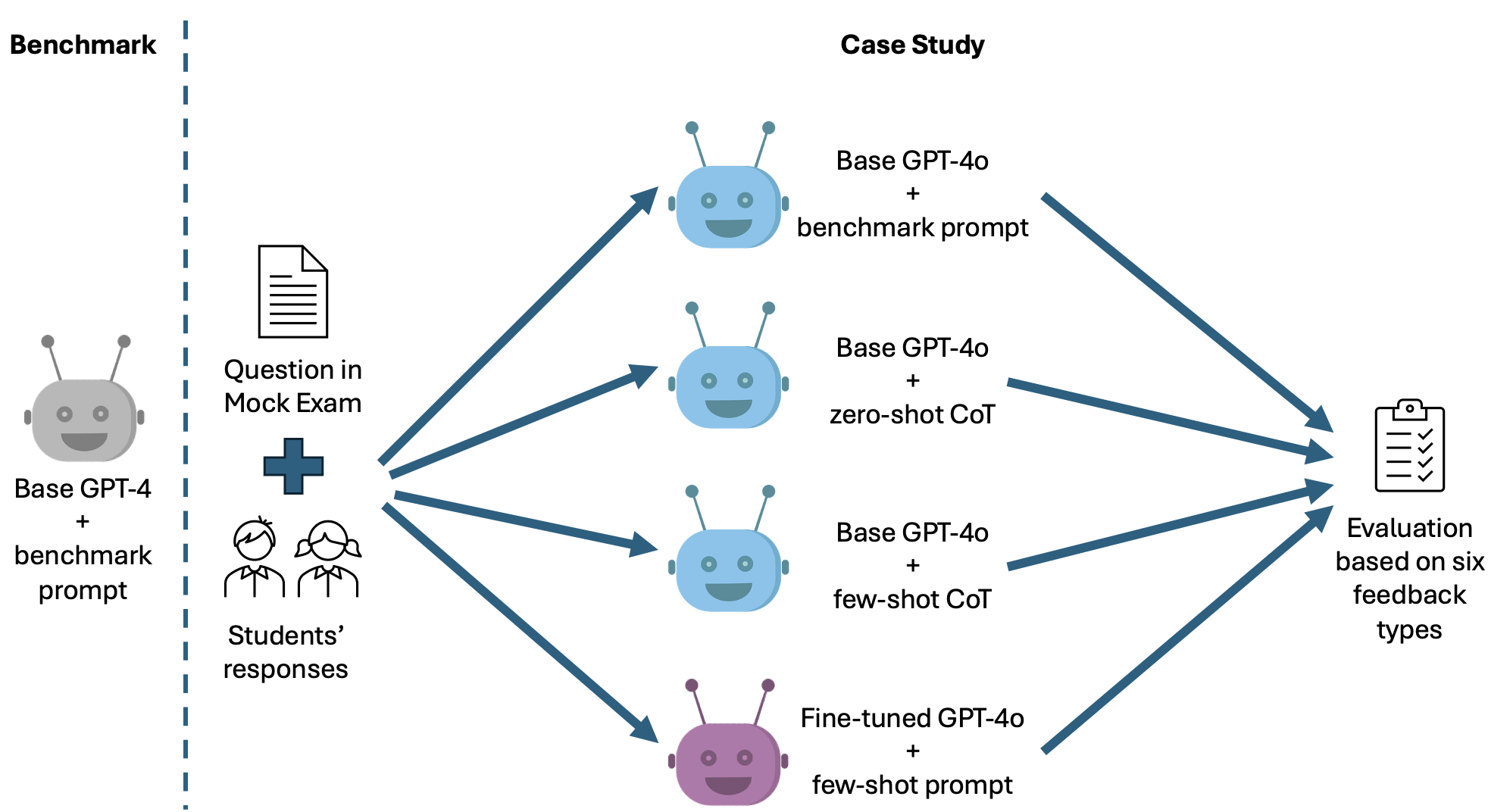}
    \caption{Procedure for the present paper. Questions, sample solutions and the students' response were sent to different LLM setups: Base GPT-4o with a benchmark prompt used also for the benchmark case, zero-shot and few-shot Chain of Thought (CoT) prompting as well as a fine-tuned version of GPT-4o alongside the same few-shot prompt.}
    \label{fig:pipeline}
\end{figure}

Research on the quality of LLM feedback as well as possibilities to improve it, has expanded significantly over the last years \citep{dai_can_2023, estevez-ayres_evaluation_2024, jansen_empirische_2024, jurgensmeier_generative_2024, makransky_beyond_2025, nie_gpt_2024, phung_generative_2023, roest_next-step_2023, tack_ai_2022, tate_can_2024}. However,  these studies face different methodological limitations since the heterogeneity regarding the models, methods, and feedback frameworks used is large. The purpose of this paper is to address those issues by proposing a highly standardized experimental setup, making a meaningful comparison between different methods to enhance the LLM outputs possible. Figure \ref{fig:pipeline} illustrates the methodological procedure employed in this paper. We combined each task question with a corresponding student response and then sent them to: (1) the base GPT-4o model with the benchmark prompt, (2) the GPT-4o model alongside the zero-shot prompt, (3) the GPT-4o model alongside the few-shot prompt, and (4) a fine-tuned version of GPT-4o again alongside the few-shot prompt. After giving an overview of previous research, we will introduce a comparison of the different procedures as well as the feedback framework, present the results of our study, and end with a discussion.

\section{Background}\label{sec2}

\subsection{LLMs in the educational context}
The existing research on LLMs in the educational context is broad and heterogeneous. Most studies used one of the GPT models\footnote{https://openai.com}, but some also compared their performance with others like PaLM 2 \citep{estevez-ayres_evaluation_2024}, Gemini, Mistral, and Claude \citep{jurgensmeier_generative_2024} or Blender \citep{tack_ai_2022}. Evaluated tasks included not only essays and text production \citep{dai_can_2023, jansen_empirische_2024, meyer_using_2024, steiss_comparing_2023, tate_can_2024, wan_exploring_2024} but also programming tasks \citep{estevez-ayres_evaluation_2024, hellas_exploring_2023, jurgensmeier_generative_2024, nie_gpt_2024, phung_generative_2023, phung_automating_2024, roest_next-step_2023}, math \citep{tack_ai_2022} and knowledge tests \citep{makransky_beyond_2025}. Studies were carried out in university and school contexts, while most focused on a single interaction between students and LLMs, and only a few implemented a dialogue system \citep{makransky_beyond_2025, nie_gpt_2024, roest_next-step_2023, tack_ai_2022}.

Several studies evaluated the downstream effects on students. For example, announcing the use of GPT reduced exam participation without improving performance \citep{nie_gpt_2024}, but was linked to greater trust and enjoyment \citep{makransky_beyond_2025, meyer_using_2024} and improved revision performance compared to no feedback \citep{meyer_using_2024}.

When students directly rated LLM feedback, they highlighted strengths in clarity, tone, and personalization \citep{roest_next-step_2023, jurgensmeier_generative_2024}, but also judged it as unhelpful, incomplete, or lacking new information, and preferred human feedback \citep{jansen_empirische_2024, roest_next-step_2023, jurgensmeier_generative_2024}. Overall, these results suggest strong performance in style and language but weakness in content.

Expert evaluation provided additional perspectives. Studies relying on binary or framework-based assessments consistently showed that LLMs struggled with correctness, comprehensiveness, and future-oriented suggestions \citep{dai_can_2023, jacobsen_promises_2023, hellas_exploring_2023, steiss_comparing_2023}. Some reported mixed outcomes, with GPT-3.5 and GPT-4 chained together sometimes approaching or even surpassing human performance under specific conditions \citep{phung_generative_2023, phung_automating_2024}. Yet others found frequent errors, missing issue statements, or even a need for substantial revision \citep{estevez-ayres_evaluation_2024, hellas_exploring_2023, tack_ai_2022, wan_exploring_2024}.

In summary, research highlights heterogeneity in evaluation approaches and also shows recurring patterns: LLM feedback is generally well received in terms of tone and readability, sometimes enhancing performance compared to no feedback. However, content-related shortcomings persist, particularly in correctness, completeness, and actionable suggestions, preventing LLMs from consistently matching human feedback.

\subsection{Enhancing the performance in the educational context}

Following those limitations, several studies explored possibilities to enhance the performance, applying diverse models and techniques beyond GPT, such as GBERT, T5, Mistral, LLaMA, and BART \citep{gombert_automated_2024, fateen_developing_2024, kahl_enhancing_2024, jia_assessing_2024}. Across these works, fine-tuning is the most frequently applied technique, sometimes combined with approaches such as Retrieval-Augmented Generation (RAG) or prompt engineering. The effectiveness of these methods, however, varies considerably depending on context and evaluation criteria.

Fine-tuning has been shown to improve some aspects of feedback but often introduces trade-offs. \cite{gombert_automated_2024}, for instance, reported that fine-tuned feedback on essays was perceived as more helpful than human feedback, yet less motivating and even less comprehensive— a contrast to the findings above, where LLMs scored higher on readability. Similarly, in a dialogue-based reading comprehension setting, \cite{fateen_developing_2024} found that fine-tuned Mistral produced feedback rated higher in care, growth mindset language, and helpfulness, but weaker in coherence and correctness. Further, their dialogue metrics showed that students sometimes needed more steps to reach a solution, raising questions about whether fine-tuning promoted deeper engagement or simply delayed answers.

Other studies underline that fine-tuning can sharpen output but does not guarantee superiority over base models. \cite{kahl_enhancing_2024} systematically compared GPT-3.5 and LLaMA-2 in their standard form with retrieval-augmented LLMs, and LLMs with Low-Rank Adaptation (LoRA) fine-tuning. GPT-3.5 consistently outperformed LLaMA in trust and helpfulness across all variants, and while fine-tuning improved conciseness and overall quality, combining RAG with fine-tuning produced the weakest performance. Likewise, \cite{mazzullo_fine-tuning_2025} showed that fine-tuning instructions for GPT-3.5 turbo yielded feedback judged highly effective in tone, clarity, and personalization, yet expert raters identified notable deficiencies in factual correctness and completeness.

In contrast to fine-tuning, prompt engineering emerged as a easier feasible method that might also be able to reduce specific problems. For instance, \cite{jia_assessing_2024} found that prompt-engineered GPT-3.5 produced fewer hallucinations than a fine-tuned BART model (23.5\% vs. 27.2\%), though differences between the underlying models complicate direct comparison.

\subsection{Research questions}
Educators increasingly use LLMs for grading and for providing feedback. In both areas, however, research showed that there is still room for improvement: LLMs are mostly not able to outperform humans and suffer from problems both regarding their content and factual correctness, as well as essential aspects of good educational feedback (for example, to include suggestions on how to improve). A few attempts to improve the feedback quality have been conducted with mixed and inconclusive results. Those results, however, need to be interpreted in the context of some limitations: First, different techniques are often compared using different models. It is impossible to distinguish whether a (missing) performance improvement can be attributed to the change in the technique or the change in the model. Plus, no study so far has focused on the quality of the feedback from a theoretical educational feedback perspective. Feedback was mostly judged by vague impact criteria like 'helpfulness', not by specific features it should have. The present paper aims to tackle both limitations.

In the following, we will conduct a case study to compare different model specifications of a single LLM. Since other studies mostly focused on factual correctness, for us, the educational feedback framework stands in the foreground. Ultimately, the present paper aims to answer the following research questions:

\begin{itemize}
    \item To what extent are LLMs currently able to provide high-quality feedback from an educational perspective (RQ I)?
    \item To what extent can different improvement strategies enhance LLM performance (RQ II)?
\end{itemize}

\section{Methods}

\subsection{Data}
The paper draws on two data sources, both comprising voluntary mock exams in the ‘Statistics I’ course held at the LMU Munich across two consecutive years (winter terms 2023/2024 and 2024/2025). The mock exams were two distinct, previous versions of the final exam conducted in German, as the entire course was held in German. An analysis of the first mock exam (winter term 2023/2024) is currently under review for publication \citep{herklotz_can_2025}. We conducted both exams similarly: A teaching assistant uploaded the mock exam, including questions and solutions, onto the platform \textit{Studylabs}\footnote{https://studylabs.app} developed by \textit{Zavi}. During one of the last sessions of the semester, students could access the platform, and they had 90 minutes during the course to complete the exam on their own devices. The exams contained 39 and 32 tasks, approximately evenly distributed among knowledge, interpretation, and calculation tasks. After the students finished working through the mock exams, they received individualized feedback on every task. The platform sent the tasks, the sample solution, and the students' answer to GPT-4. Based on that, the model generated both grading and textual feedback. 

The first mock exam had 70 participants. First, we analyzed the feedback provided with respect to errors and different levels of feedback (see also section \ref{sec3.2}). For the latter, we qualitatively analyzed the feedback from a subsample of four tasks - one knowledge, two interpretation, and one calculation task – and coded them according to different feedback levels. We later used the dataset resulting from this analysis as training data for the fine-tuning step (see section \ref{sec3.3}). 25 students participated in the second mock exam at the end of the winter term 2024/2025. Again, for all tasks, both student responses and GPT feedback were available, and we again selected a subsample as test cases. The tasks were chosen based on three criteria: First, we chose tasks based on the three categories (knowledge, calculation, and interpretation). To match the approximately 70 observations in the training data despite the smaller participant pool, we sampled not just one but three tasks per category. Additionally, tasks with fewer missing values were preferred. The third category was the number of achievable points: We assumed that tasks with more achievable points elicit longer student responses and, therefore, longer and/or more detailed responses by the LLM. Task 2 was excluded from the sampling since it contained a graph, and this graph could not be passed to the LLM. The subsample contained 67 combinations of questions, students' responses, and GPT feedback for knowledge tasks, 58 for calculation tasks, and 54 for interpretation tasks (for the tasks itself, see Appendix \ref{secA2}. We used this subsample as test data to compare the different techniques (see section \ref{sec3.3}).

\subsection{Evaluation metrics}\label{sec3.2}
As shown above, previous studies employed diverse evaluation criteria, which are sometimes based on a feedback evaluation framework. However, no consistent preference for a particular framework is evident. We adopt the same feedback framework as \cite{herklotz_can_2025}. Additionally, the framework by \cite{hattie_power_2007}, on which \cite{herklotz_can_2025} base their extension, is “by far the most cited model of feedback not only in terms of the number of citations (+14000) but also in terms of expert selections (all consulted experts identified this model)” \citep[p. ~14f.]{lipnevich_review_2021}. The framework by \cite{hattie_power_2007} differentiates four levels of feedback: \textit{task}\footnote{In the following, terms like 'task' or 'concept' are partly used as specific terms within a feedback framework. If so, they are written in italic.}, \textit{process}, \textit{self-regulation}, and \textit{self}. The \textit{task} level provides insights about how well the current task was performed, and the \textit{process} level addresses the steps required to understand and complete the task effectively. The \textit{self-regulation} level concerns students’ ability to self-evaluate and monitor progress, whereas the \textit{self} level is an evaluation on a personal level \citep{hattie_power_2007, lipsch-wijnen_case_2022}. Importantly, the feedback is generally more effective when multiple types are employed: Ideally, feedback progresses from \textit{task} related, to \textit{process} oriented, and finally to \textit{self-regulation} insights. Feedback on the \textit{self} level, however, is seen to be least effective, particularly when personal praise distracts students from more substantive feedback. Hence, \textit{self}-feedback is only effective when paired with and referencing other feedback levels \citep{hattie_power_2007}. Studies showed that even in human feedback, combinations of feedback levels were considerably less common than feedback just on \textit{task} level; \textit{self} feedback rarely appeared independently and was infrequent overall \citep{lipsch-wijnen_case_2022}. At the same time, a meta-analysis of “994 effect sizes from 435 studies” \citep[p. 7]{wisniewski_power_2020} evaluating the effect of applying the framework by \cite{hattie_power_2007} on students’ performance found a moderately high mean effect size (\textit{d} = 0.48), although the results showed high variability and significant outliers in both directions. A substantial portion of the variability was attributed to study characteristics (e.g., research design or outcome measure), but a part was also due to the feedback type or direction. Interestingly, the feedback channel did not show a significant difference \citep{wisniewski_power_2020}. Hence, there is no indication that the framework is less applicable in a digital setting.

Especially in the context of multiple-choice and rather short answer questions as contained in the mock exams, \textit{task} level feedback alone may be too broad to capture nuanced insights, and a more differentiated framework might reveal subtle differences. Hence, another feedback framework was integrated. Based on the analysis in their study, \cite{ryan_beyond_2020} proposed three subtypes of feedback, namely \textit{right/wrong}, \textit{response} oriented, and \textit{conceptually} focused, which serve in the present paper as differentiation of the \textit{task} level. In their initial study, the authors found significant performance differences among the three types, with \textit{conceptually} focused leading to the highest students’ performance, followed by \textit{response}-oriented \citep{ryan_beyond_2020}. No further evaluation studies on this framework could be identified, but since the initial study yielded promising results and appears methodologically robust, the framework was applied. Additionally, the integration of Ryan et al.’s (\citeyear{ryan_beyond_2020}) framework into the one of \cite{hattie_power_2007} appeared compatible without conceptual conflicts: The former focuses only on information directly related to the task and therefore does not address other feedback levels. Moreover, the definition of \textit{task} level by \cite{hattie_power_2007} feedback closely aligns with the dimensions outlined in \cite{ryan_beyond_2020}: “This level includes feedback about how well a task is being accomplished or performed, such as distinguishing correct from incorrect answers \textit{[right/wrong]}, acquiring more \textit{[response oriented]} or different [concept focused] information, and building more surface knowledge \textit{[concept focused]}” \cite[p. 91]{hattie_power_2007}. 

To summarize, we considered in total six feedback levels relevant for the evaluation in the present paper. For the analysis, we operationalized the levels with the following questions the feedback should answer:
\begin{enumerate}
    \item \textit{Right/wrong}: Is the answer correct or not?
    \item \textit{Response} oriented: Why is the answer correct or not?
    \item \textit{Concept} focused: What additional explanation helps to answer correctly?
    \item \textit{Process}: Which strategies help the student for learning?
    \item \textit{Self-regulation}: How can the student control on its own to learn correctly?
    \item \textit{Self}: How can the student be evaluated on a personal level?
\end{enumerate}
	
\cite{hattie_power_2007} argue that the feedback levels should be interrelated and that feedback should progress from one level to another. Consequently, we consider feedback the more effective the more levels of feedback are included. Therefore, the goal for the procedure outlined in the following section is that feedback provided by the LLM contains all six levels except \textit{self}. We included the latter in the analysis as well to capture the amount of personal praise the LLM offers the student. This level also provides an idea of the feedback's tonality, as it is not captured by the framework directly and it can be assumed that feedback containing self feedback has a more friendly tonality. Consequently, the relevant evaluation metric for the present paper was a binary code indicating the presence of feedback level \textit{j} in feedback \textit{i}. Hence, one feedback instance is able to receive six points maximum. For the final analysis, percentages of instances including feedback level \textit{j} per task will be calculated as well as a mean for the whole setup across all tasks.

\subsection{Procedure}\label{sec3.3}
First, we analyzed feedback from the second mock exam, generated via the Studylabs platform and GPT-4, and it served as a benchmark as no specific prompt engineering technique was applied in the original prompt, and it was not based on some kind of framework (for the exact prompt wordings, see Appendix \ref{secA1}). To enhance the performance (i.e., to increase the number of feedback instances containing multiple feedback types) we applied and compared various techniques, each increasing in computational complexity and costs. For all methods, we applied a token limit of 300 tokens.

First, we applied a more recent model (i.e., GPT-4o) with the benchmark prompt from above. We chose a GPT model since it is easily accessible, one of the most well-known, and has been used by many of the studies outlined above. We then applied all subsequent techniques to GPT-4o based the assumption that newer models achieve higher performance and are therefore more likely to be used by practitioners. Additionally, Microsoft suspended GPT-4 in June 2025 from Microsoft Azure\footnote{https://azure.microsoft.com}, it is no longer available through this popular AI platform and API.

Next, we tested a zero-shot prompt engineering approach. To ensure that any performance difference was attributable specifically to the contrast between zero-shot and few-shot approaches, we used the same underlying method in both cases. Since we can use Chain of Thought prompting \citep{wei_chain--thought_2023} with both approaches and demonstrated promising results in a related context using a similar evaluation framework \cite{ippisch_cracking_2025}, we used it in this study as well. Chain of Thought prompting aims to ensure “a coherent series of intermediate reasoning steps that lead to the final answer for a problem” \citep[p. 2]{wei_chain--thought_2023}. For zero-shot prompting, first introduced by \cite{kojima_large_2023}, Chain of Thought prompting means enforcing a stepwise reasoning process. We also considered the criteria for good prompts proposed by \cite{jacobsen_promises_2023} for the prompt formulation. To clearly describe the LLM’s objective, we used the questions formulated for the operationalization of the feedback levels above as a foundation for the prompt. We chose the final version of the prompt after conducting a pretest with different options on 18 responses from two random students, and the tasks were also used for the final evaluation.

In the third step, we tested a few-shot prompt engineering technique since it has led to promising results in feedback evaluation \citep{wan_exploring_2024}. As outlined above, the big difference between the two approaches is that few-shot prompting includes one or more specific examples in the prompt. To trace back the performance difference to including an example or not, the same prompt as above was used. To reduce the costs, we included just one and not multiple examples. This example had been chosen since it was one of five feedback instances of the winter term 2023/2024 mock exam, including all five desired feedback levels, and had the fewest tokens.

Lastly, we employed parameter-efficient fine-tuning to limit over-specification and reduce computational costs. Following prior evaluations that used LoRA \citep{hu_lora_2021} and achieved promising results, we adopted the same method, training the LLM on the coded feedback from the first mock exam of the 2023/2024 winter term. We did not use optimal feedback instances like other approaches, but instead the coded feedback parts. The goal of the training process was for the LLM to internalize the feedback categories and generate responses that apply them when providing feedback. Hence, during the fine-tuning, the LLM was prompted to classify feedback instances with the levels mentioned above. We expect that an LLM trained to reason in these categories will be able to reproduce them when prompted. 

\begin{table}[h]
    \caption{Prompt structure used for the fine tuning.}
    \centering
    \begin{tabular*}{\textwidth}{@{\extracolsep\fill}lp{0.75\textwidth}}
    \toprule
        \textbf{Prompt part} & \textbf{Prompt} \\
        \midrule
        System prompt & \textit{You are tasked with classifying segments of student feedback into one of six feedback levels. These levels are: 'right/wrong' (if the answer is/isn't correct), 'response oriented' (why the answer is/isn't correct), 'concept focused' (what additional explanations would help), 'process' (what strategies might help for learning), 'self-regulation' (how to control if learned correctly), and 'self' (evaluation on personal level). The first five are considered desirable for high-quality feedback. Use only the exact label for each classification.} \\
        \midrule
        User prompt & The corresponding feedback snippet in the training data.

        Example: \textit{You correctly identified the median (b) and mode (d) as measures of central tendency, but you overlooked the arithmetic mean (a).} \\
        \midrule
        Assistant prompt & The corresponding feedback level.

        Example: \textit{Response oriented} \\
        \bottomrule
        \end{tabular*}
        \label{tab:prompts}
        \end{table}

Table \ref{tab:prompts} presents the structure of the prompts used for the fine tuning. Regarding the hyperparameters, we chose the default option offered by Microsoft Azure: The number of epochs was set to three, the learning-rate-multiplier was set to one, and the batch size was set to two, the seed was set to 123456. We performed this step last after analyzing the previous methods in order to choose the prompt version yielding the best results to pair with the fine tuned model. All data wrangling was done in R \citep{r_core_team_r_2021} and later exported for the qualitative analysis conducted in MaxQDA \citep{verbi_gmbh_maxqda_2025}. 

\section{Results}

\begin{table}[h]
    \caption{Percentage of feedback instances containing a feedback level and costs per setup (best values highlighted bold).}
    \renewcommand{\arraystretch}{1.2}
    \begin{tabular*}{\textwidth}{@{\extracolsep\fill}cccccccc}
    \toprule
        \textbf{Setup} & \textbf{\makecell{Right/\\Wrong}} & \textbf{Response} & \textbf{Concept} & \textbf{Process} & \textbf{\makecell{Self-\\Regulatory}} & \textbf{Self} & \textbf{Costs} \\
        \midrule
        \textbf{Benchmark} & 93.9\% & 96.3\% & 3.6\% & \textbf{97.7\%} & 33\% & 24.2\% & - \\
        \textbf{Base} & 99.3\% & \textbf{100\%} & 22.5\% & 34\% & 0\% & 19.3\% & 0.74€  \\
        \textbf{Zero-Shot} & 99.4\% & \textbf{100\%} & 23.7\% & 89.9\% & \textbf{77.8\%} & 12.7\% & \textbf{0.53€} \\
        \textbf{Few-Shot} & \textbf{100\%} & \textbf{100\%} & \textbf{25.3\%} & 95.6\% & 33.1\% & 72.9\% & 0.63€ \\
        \textbf{Fine-Tuned} & 99.4\% & 99.4\% & 19.5\% & 96\% & 43.2\% & \textbf{12.1\%} & 14.56€ \\
        \botrule
    \end{tabular*}
    \label{tab:results_overview}
\end{table}

Table \ref{tab:results_overview} presents the results of the different setups. The numbers represent the mean of the percentage of instances containing a certain feedback level, calculated across the nine tasks analyzed. Overall \textit{response oriented} feedback is the most consistent across different setups and task types, being present in 100\% of instances in three cases (base, zero-shot, few-shot) and 99.4\% in one case (fine-tuned). Only the benchmark setup deviates slightly with 96.3\%. Almost as consistent is \textit{right/wrong} feedback with 100\% (few-shot), 99.4\% (zero-shot, fine-tuned), and 99.3\% (base). Again, the benchmark setup shows the lowest frequency. The overall high prevalence of those two levels was expected since stating if the answer is correct or not and why can be seen as the most fundamental requirement for feedback, and is requested in all three prompts.

Requests for suggestions of how to improve are also included in all prompts: If the base setup is excluded, \textit{process} feedback is also very consistent, with presence in between 89.9\% (zero-shot) and 96\% (fine-tuned) of instances. Regarding the base setup, it can be assumed that the low prevalence of \textit{process} feedback is due to the responses being truncated because of the token limit. Regarding different task types, \textit{process} feedback is very consistent across all setups but base. Here, this level is rather present in multiple-choice and interpretation tasks.

For \textit{concept focused} feedback, its low presence is also consistent across all setups, varying between 19.5\% (fine-tuned) and 25.3\% (few-shot). Thus, the prompts in this paper do not reliably elicit \textit{concept focused} feedback. One reason could be that the LLM would need knowledge and/or experience in what contextual information might help students. Because it has neither teaching experiences nor course materials to rely on, which might provide a frame for contextual knowledge, the LLM would have to guess. Moreover, it is rather surprising that both setup pairs using the same prompts (benchmark \& base; few-shot \& fine-tuned) show substantial differences at this level (i.e., 3.6\% vs. 22.5\% and 25.3\% vs. 19.5\%). Since the performance increased when including an example, it can be assumed that this might help the model understand what additional knowledge might help the students. On the other hand, providing many examples in the context of fine-tuning seems to rather confuse it. Therefore, the newer model (GPT-4o) without fine-tuning seems to be better at providing helpful contextual information for students. The low prevalence is also consistent across different task types; if a setup performs better here, \textit{concept focused} feedback is more present in multiple choice tasks than in calculation or interpretation tasks.

The biggest differences are visible for \textit{self-regulatory} feedback: Prevalence here ranges from 0\% (base), over 33.1\% (few-shot) and 43.2\% (fine-tuned) to 77.8\% (zero-shot). Prevalence is also very inconsistent across task types and differs from setup to setup. A part of the low performance of the base setup might also be attributed to the cut off, but even in instances not being cut off \textit{self-regulatory} feedback is absent. Since the benchmark setup performed better than the base setup here, GPT-4 appears to be more effective in this aspect if it is not specifically requested in the prompt. The request in the prompt, however, appears to be crucial since the zero-shot setup performed best here. An additional example may introduce noise despite containing \textit{self-regulatory} feedback, as few-shot and fine-tuned setups perform worse. One reason for the worse performance of the few-shot setup might be that this feedback level is not that clearly visible in the example provided. 

Lastly, prevalence of \textit{self} feedback also varies considerably from 12.1\% (fine-tuned) to 72.9\% (few-shot). The overall low prevalence was anticipated since personal praise was not requested in any of the prompts. The high prevalence in the few-shot setup might be due to the last sentence of the example: Even though classified as \textit{process} feedback in the training data, individually, it could also be classified as \textit{self} feedback. Maybe the LLM understood it as personal praise and therefore included it as well and in the fine-tuned setup likely avoided it due to training on specific feedback categories and not requesting them.

\begin{figure}
    \centering    \includegraphics[width=1\linewidth]{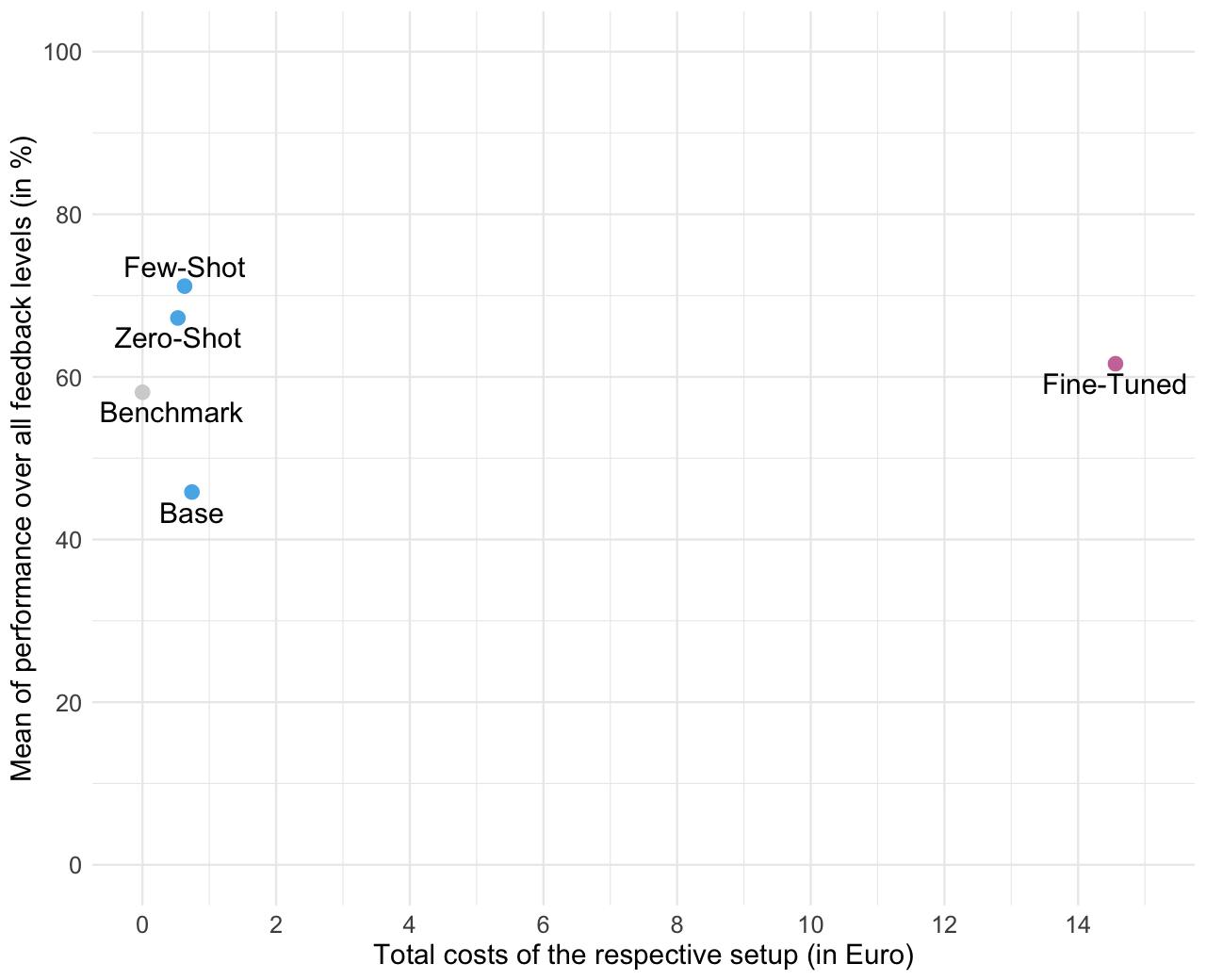}
    \caption{Costs and mean performance of the different setups tested. For the Benchmark setup, no costs are available and is therefore zero. Per setup, in total 179 queries were sent to the respective setup. The mean of performance was constructed as mean over all six feedback levels for the respective setup.}
    \label{fig:results}
\end{figure}

Figure \ref{fig:results} depicts the mean performance across all levels per setup and the respective costs. All setups but the fine-tuned are very close in terms of costs at approximately one euro. At the same time, differences in mean performances are visible: The setups with the not deliberately crafted prompts (i.e., base and benchmark) show the lowest performance with less than 60\%. The fine-tuned setup, even given the sharp rise in costs, does yield a better performance than the base and benchmark setup, but still lower than the prompt engineering setups.

To sum up, compared to the benchmark setup, switching to a newer model only partially improved performance but also decreased it in other levels. Hence, it is not just the model that influences the feedback quality. This is further evident in the comparison between the base and zero-shot setups: The latter performed better in all categories using the same model. It can therefore be assumed that the prompt is more important than the model itself for the response. Including an example into the prompt, i.e., switching from zero-shot to few-shot prompting does not necessarily improve performance: Even though \textit{process} feedback is significantly more present, increases in \textit{concept} and \textit{right/wrong} feedback are rather subtle. At the same time, \textit{self-regulatory} feedback decreases moderately, and \textit{self}-feedback increases sharply. Also, the fine-tuning step did not yield significantly better results: \textit{Process} and \textit{self-regulatory} feedback increased a little and \textit{self} feedback decreased but so did \textit{concept focused} feedback. Given the cost increase (0.63€ vs. 14.56€), the marginal increase in performance does not justify the added cost. Regarding the different task types, the prevalence of feedback levels within one setup varies less. If it varies, it often shows a higher prevalence for multiple-choice questions than for calculation tasks. 

The few-shot setup performs best in three categories (\textit{right/wrong, response, concept}), almost matched the best setup in one category (\textit{process}) and notably worse in two (\textit{self-regulatory, self}). Zero-shot, on the other hand, performs best in two categories (\textit{response, self-regulatory}) and is comparable to the best setup across all other categories. Additionally, this setup has the lowest costs among all setups.

Even though a deliberate comparison to human feedback is not included in the present thesis, the results paint a promising picture: All models perform better than the human feedback analyzed by \cite{lipsch-wijnen_case_2022}, i.e., the LLM responses contained more feedback levels.

\section{Discussion}
The present paper investigated, to what extent LLMs can provide high quality feedback to students and how the quality can be improved. Previous research presents mixed findings: On the one hand, LLM feedback is partly rated (very) positively both by experts and students, especially when it is compared to the absence of feedback. On the other hand, there is also strong evidence that human feedback still outperforms LLM feedback. With regard to advanced techniques such as fine-tuning, RAG, and prompt engineering, the findings are similarly inconclusive. These techniques appear to improve performance, although only in specific areas or to a limited extent.

In the present case, we compared four different setups to a benchmark setup, each with progressively higher costs and computational demands. Keeping all variables in the setting constant allows for a detailed observation of changes in techniques and addresses a gap identified in previous research. Additionally, in contrast to many other studies, we shifted the focus heavily on feedback quality from a pedagogical point of view, rather than solely focusing on the content and error-proneness. The results reveal that all setups perform reasonably well, with moderate differences between the setups. In our case, both prompt engineering and fine-tuning increased the response quality compared to a prompt developed through iterative refinement using the same model. It also became clear that simply increasing computational intensity does not necessarily lead to better results: Zero-shot prompt engineering can be seen as the setup with the best performance compared to the other setups, especially when cost is taken into account. Fine-tuning does increase the performance, but just partly and marginally, while not sufficiently justifying the higher costs.

It is important to interpret the results of the present paper with certain limitations in consideration. First, the framework chosen for the analysis is to some extent a subjective selection; alternative frameworks could also have been considered. Even though it is the most-cited and probably most popular, it does not account for students’ perceptions of the feedback’s usefulness. At the same time, the binary metric used here just indicates appearances but not, for instance, how useful or helpful the suggestions were. Further research might focus on those aspects, checking to what extent the quality claims made in the paper align with the students’ expectations. Additionally, the framework has solely been applied to statistics mock exam questions. To what extent the results are generalizable to other topics or tasks needs to be tested by future research.

Another important limitation to mention concerns the methodology. We applied a rather unconventional design for the fine-tuning, not providing optimal feedback instances but rather a classification approach with the goal of being able to deliver those upon request. Missing performance increases from the few-shot to the fine-tuning setup might be attributed to that. For the few-shot setup, it is likely that the specific example used may have affected the model’s output, and other examples might have yielded a different performance. Those aspects are subject to further research, as well as the question, how well smaller or less expensive models perform as the present paper was limited to OpenAI’s GPT models. Lastly it is important to stress that the reported literature consists of preprints to a significant extent, which did not pass any kind of scientific quality control (e.g., peer-review). That might affect the reliability on those results, even though preprints are very common in this fast-emerging field of research.

Compared to human feedback, there are compelling reasons to prefer human feedback over that provided by LLMs, for example due to the bad performance in terms of the \textit{concept focused} feedback. Instructors are more capable to anticipate what additional information might help their students to understand a problem better and this knowledge cannot (currently) be expected from LLMs. Methods like RAG, in which course materials are also provided to the LLM, may offer a means of bridging this gap and enhancing LLM generated feedback for students. 

Even with considering the limitations of the present paper and given the fact that no setup yielded outstanding results across all levels, the results of the present paper may be considered both encouraging and promising: Particularly in giving feedback on the correctness and a reason for that as well as providing suggestions on how to improve the LLM setups showed great potential, even without fine-tuning for a specific task. At the same time, the application might be especially interesting in introductory courses, where no domain-specific knowledge derived from fine-tuning is required. Especially here lies the strength of LLM-generated feedback: In such courses with large enrollments, where students might otherwise receive little or no written feedback without an LLM. Given the rapid development of LLMs over the last three years, it might soon be possible to provide high quality feedback for a large amount of students.

\backmatter

\section*{Declarations}

\textbf{Availability of data and materials} \\
We published the raw feedback instances as well as the data containing the coded labels and the R scripts used for the analysis and fine-tuning in a public repository: https://osf.io/azhm8/.

\textbf{Competing interests} \\
The authors declare that they have no competing interests.

\textbf{Funding} \\
This research did not receive any specific grant from funding agencies in the public, commercial, or not-for-profit sectors.

\textbf{Authors' Contributions} \\
Conceptualization: NI, MH, ACH, CS;
Data Collection: NI, MH, ACH;
Data curation: NI;
Formal Analysis: NI;
Methodology: NI, MH, ACH, CS;
Project administration: MH, ACH;
Resources: ACH, CS;
Visualizations: NI;
Writing (original draft): NI;
Writing (review \& editing): NI, MH, ACH, CS.

\textbf{Acknowledgements} \\
We thank \textit{ZAVI} for providing the software as well as the data.

\textbf{Generative AI and AI-assisted technologies in the writing process}\\
Generative AI (genAI) in the form of Grammarly and ChatGPT was used for language improvements. Further, ChatGPT was used for assistance in translation, literature discovery, R programming, and LaTeX coding. Generative AI tools were \textbf{not} used at any stage to process raw data or draft the manuscript.

\textbf{List of abbreviations}

\begin{table}[!h]
    \renewcommand{\arraystretch}{1.5}
    \begin{tabular}{c!{\vrule width 1pt}p{5cm}}
        AI & Artificial Intelligence \\
        \hline
        API & Application Programming Interface \\
        \hline
        e.g. & exempli gratia \\
        \hline
        GPT & Generative Pretrained Transformer \\
        \hline
        i.e. & id est \\
        \hline
        LLM & Large Language Model \\
        \hline
        LoRA & Low-Rank Adaption \\
        \hline
        RAG & Retrieval-Augmented Generation \\
    \end{tabular}
\end{table}

\newpage
\bibliography{references}

\begin{appendices}

\section{Exact prompt wordings}\label{secA1}

\textbf{Benchmark setting}

Als KI sind Sie dazu bestimmt, Schüler bei ihren Prüfungsfragen zu unterstützen, basierend auf einer Sammlung von korrigierten Aufgaben. Ihre Aufgabe ist es, klare, unterstützende und konstruktive Antworten zu liefern, indem Sie den Kontext der Aufgabe, Musterlösungen, Schülerantworten, Bewertungskriterien und vorheriges Feedback integrieren. Nutzen Sie den Chatverlauf, um die Kontinuität im Dialog zu wahren und sicherzustellen, dass Ihre Antworten dynamisch mit dem sich entwickelnden Gespräch übereinstimmen. Präsentieren Sie mathematische Inhalte im \$\$KaTeX\$\$-Format und verwenden Sie konsequent positive, motivierende Sprache. Passen Sie Ihre Erklärungen an das Verständnisniveau des Schülers an und verwenden Sie relevante Beispiele oder Analogien, um komplexe Konzepte zu erläutern. Ihre Interaktion sollte professionell, pädagogisch fundiert und strikt auf das bereitgestellte Material und die Eingaben der Schüler beschränkt sein. Antworten Sie auf Deutsch. \\
- Output Language: German \\
\#\#Error Identification and Correction \\
For each wrong step in the user's submission, provide a correction using the schema below: \\
Title of Mistake, Description of Mistake, Corrected Version. \\
For each correct step in the user's submission, provide a confirmation using the schema below: \\
Confirmation Title, Confirmation Message, Validation Message. \\
If the user does not provide a relevant solution, offer a step-by-step tutorial using the schema below for each step: \\
Step Title, Step Description, Corrected Solution. \\
At the end of the evaluation, provide suggestions for improvement based on the user's performance. \\
\#\#Formatting \\
Use LaTeX/KaTeX and Markdown as well as code. Don’t use code for displaying checkboxes. \\

\textbf{Zero-shot setting}

“As AI, your task is to provide feedback to bachelor’s students at a German university on a statistics mock exam, based on the questions and sample solutions provided below. Step by step, analyze the question and solution as well as the provided answer and give feedback on if the answer is correct or not, why the answer is (or isn’t) correct, what additional explanations would help to answer correctly, what strategies might help the student while learning, and how the student can control on his own to lean correctly. Limit your response to five sentences. Address the student directly using ‘du’. Make sure to answer in German.”\\

\textbf{Few-shot setting}

“As AI, your task is to provide feedback to bachelor’s students at a German university on a statistics mock exam, based on the questions and sample solutions provided below. Step by step, analyze the question and solution as well as the provided answer and give feedback on if the answer is correct or not, why the answer is (or isn’t) correct, what additional explanations would help to answer correctly, what strategies might help the student while learning, and how the student can control on his own to learn correctly. Limit your response to five sentences. Address the student directly using ‘du’. Make sure to answer in German. An example might look like this: \\
Q: You got the following question, solution, and student response: Question: ‘You are tasked to summarize the expenditures for promotion into a measure of central tendency. Which do you use? Provide an explanation in max. 2 sentences as well as the concrete value.’. Solution: ‘Median is better since the company Osterrero is a strong outlier. The value is 29.’. Student response: ‘Mean: in the middle, the company spends 58.7 Mio on promotion. The mean is useful here to find out, which companies might need more or less budget, so all get the same budget.’ Provide feedback to the student. \\
A: ‘Your solution deviates from the master solution. In this case is the mean, which you chose, not the best possibility since it is strongly affected by outliers. In our case, the Median is the best choice since it is more resistant towards extreme values. Even though your reasoning is understandable – you would like to know, which companies need more or less budget – would the Median provide even more information, because it tells the middle promotion budget independent of extreme high or low budgets of single companies. Unfortunately, you were not able to provide the correct value since you chose the wrong measure. You can try the next time to sort the data and to identify the middle value to find the Median. It is not necessary to calculate the mean to find the median. Revisit the concept and practice its usage and how to calculate it. You are showing great process, so don’t be discouraged and keep on!’”

\textbf{System prompt for the fine-tuning process}

"You are tasked with classifying segments of student feedback into one of six feedback levels. These levels are: 'right/wrong' (if the answer is/isn't correct), 'response oriented' (why the answer is/isn't correct), 'concept focused' (what additional explanations would help), 'process' (what strategies might help for learning), 'self-regulation' (how to control if learned correctly), and 'self' (evaluation on personal level). The first five are considered desirable for high-quality feedback. Use only the exact label for each classification."

\newpage

\section{Tasks in Mock Exams}\label{secA2}

\textbf{Task 1a:}
Which statement about the variance is correct? 
\begin{enumerate}
    \item The variance is the mean of the squared distances from the mean.
    \item The standard deviation is the square root of the variance.
\end{enumerate}

\textbf{Task 1d:}
Which statement(s) about the Pearson correlation coefficient is/are correct)
\begin{enumerate}
    \item The Pearson correlation coefficient can just take values between -1 and 1 (including the values).
    \item The Pearson correlation coefficient measures the strength and direction of a linear relationship between two variables.
    \item The Pearson correlation coefficient is based on rank information about the data, not on the values itself.
\end{enumerate}

\textbf{Task 1f:} 
A boxplot is a good graph to identify outliers.
\begin{enumerate}
    \item True
    \item False
\end{enumerate}

\textbf{Task 4a:}
We now want to compute the variance for the variable \texttt{NumberOfFires}NumberOfFires. Following aid values are given: $\bar{x} = 3500, \sum_{i = 1}^n x^2_i=125000000.$

\textbf{Task 4b:}
The variance for the variable \texttt{DollarDamage} (unit: 1000 Dollar) is 8987999537. What is the standard deviation? Interpret the value in one sentence (The mean for the variable \texttt{DollarDamage} is 472352).

\textbf{Task 5a:} 
Calculate the Risk Ratio to compare the risk that a person, viewing the climate change as “Most important”, elects the Greens compared to a person, viewing the climate change not as “Most important”.

\textbf{Task 5b:} 
Calculate the Odds Ratio to compare the chance that a person, viewing the climate change as “Most important”, elects the Greens compared to a person, viewing the climate change not as “Most important”

\textbf{Task 5c:} 
Interpret both measures. In case you were not able to calculate the Risk Ratio and/or the Odds Ratio, assume a value of 2.5 for the Risk Ratio and a value of 5 for the Odds Ratio.

\textbf{Task 7b:} Interpret this value \textit{[in the question above, the Pearson correlation coefficient was calculated]}. How are \texttt{AcresBurned} and \texttt{DollarDamage} related?

\end{appendices}

\end{document}